\documentclass[12pt]{article}
\usepackage{graphicx,times}
\usepackage{epsf}
\usepackage{amsmath,amssymb,latexsym,float}
\usepackage{enumerate}
\usepackage{listings}
\numberwithin{equation}{section}

\newcommand{\R}{\text{\fontshape{n}\selectfont I\kern-.42exR}}

\newcommand{\1}{\text{\fontshape{n}\selectfont 1\kern-.56exl}}

\begin{document}
\title{
{\bf On Gauged Renormalisation Group Transformations of Lattice Fermions}
}

\author{Artan Bori\c{c}i\\
        {\normalsize\it University of Tirana}\\
        {\normalsize\it Department of Physics, Faculty of Natural Sciences}\\
        {\normalsize\it King Zog I Boulevard, Tirana, Albania}\\
        {\normalsize\it borici@fshn.edu.al}\\
}

\date{}
\maketitle

\vspace{2cm}
\begin{abstract}
We construct a hierarchy of lattice fermions, where the coarser lattice Dirac operator is the Schur complement of the block UL decomposition of the finer lattice operator. We show that the construction is an exact gauged renormalisation group transformation of the lattice action. In particular, using such a transformation and the QCDLAB tool, it is shown how to implement the Ginsparg-Wilson strategy for chiral fermions in the presence of a dynamical gauge field. The scheme allows, for the first time, a full multigrid algorithm for lattice quarks.
\end{abstract}

\vspace{14cm}
\pagebreak


{\bf 1.} A renormalisation group transformation for quadratic actions, as it is the case of lattice fermions, is a simple Gaussian integration,
$$
\det\tilde D~e^{-\bar\psi_b S_{bb} \psi_b}=\int_{\bar\phi\phi}e^{-(\bar\psi_b-\bar\phi\bar B)D_{bb}(\psi_b-B\phi)-\bar\phi D\phi}\ ,
$$
where $D$ and $S_{bb}$ are Dirac operators on the fine and coarse lattices, $D_{bb}$ is a naive Dirac operator on the coarse lattice, $B,\bar B$ are blocking operators, and, by evaluating the right hand side, one can show that
$$
\tilde D=D+\bar B D_{bb}B, ~~~~~~~S_{bb}=D_{bb}-D_{bb}B{\tilde D}^{-1}\bar BD_{bb}\ .
$$

This general transformation can be made concrete by the following prescription:

\noindent
{\it 1. Partition the Dirac operator as a 2x2 block operator in the form},
$$
D=
\begin{pmatrix}
D_{bb} & D_{br}\\
D_{rb} & D_{rr}
\end{pmatrix}
$$
{\it 2. Define the gauge covariant blocking kernels},
$$
B=
\begin{pmatrix}
0 & D_{bb}^{-1}D_{br}
\end{pmatrix}, ~~~~~\bar B=
\begin{pmatrix}
0\\
D_{rb}D_{bb}^{-1}
\end{pmatrix}\ .
$$
In this case, one can show that:
$$
\det\tilde D=\det D_{bb}\det D_{rr}, ~~~~~~S_{bb}=D_{bb}-D_{br}D_{rr}^{-1}D_{rb}\ .
$$

On the other hand, a block UL decomposition of $D$ gives
$$
\begin{pmatrix}
D_{bb} & D_{br}\\
D_{rb} & D_{rr}
\end{pmatrix}
=
\begin{pmatrix}
I_{bb} & D_{br}D_{rr}^{-1}\\
0 & I_{rr}
\end{pmatrix}
\begin{pmatrix}
S_{bb} & 0\\
D_{rb} & D_{rr}
\end{pmatrix}\ ,
$$
$S_{bb}$ being the {\it Schur complement}. Therefore, we may conclude that:
\begin{itemize}
\item[i)  ] the coarse lattice Dirac operator is the Schur complement $S_{bb}$ of the block UL decomposition of the fine lattice Dirac operator;
\item[ii) ] the fermion measure of the original theory is $\det D=\det S_{bb}\det D_{rr}$;
\item[iii)] if $D$ is the continuum Dirac operator, the Ginsparg-Wilson relation reads \cite{GW}:
$$
\{\gamma_5,S_{bb}^{-1}\}=\{\gamma_5,D_{bb}^{-1}\}\ .
$$
\end{itemize}

{\bf 2.} In their paper, Ginsparg and Wilson note that {\it ``any $h$, in particular a fixed point $h$ approached after many iterations of the block spin transformation is said to be chirally invariant if it satisfies''} the relation.
\footnote{In their notations the relation is $\{\gamma_5,h^{-1}\}=\{\gamma_5,\alpha^{-1}\}$.
} They conclude the paper by noting: {\it ``Finding a way to go ahead and actually gauge a symmetry present only in remnant form stands as a further challenge''}.

Since then, a lot of efforts have been devoted to solve the Ginsparg-Wilson relation: the overlap fermion is an exact solution to this relation \cite{NaNe93,Ne98}, whereas domain wall fermions \cite{Ka92,FuSha95} and classical fixed point actions \cite{Hasenfr_et_al98} lead to approximate solutions.

Since a fixed point approach in the presence of gauge fields will be a very complicated function of the original fine lattice operator, we propose a hybrid approach: {\it trading the Dirac kernel of the one step gauged renormalisation group, or possibly its approximation, for the kernel of the overlap or domain wall fermions}. With one step only, we retain simplicity and hope to go away from the Aoki phase. In that case, we would get domain wall fermions with smaller residual mass and overlap fermions with better localisation properties.

\begin{figure}
\vspace{5.5cm}
{\hspace{-4cm}\epsfxsize=8cm \epsffile[10 20 300 180]{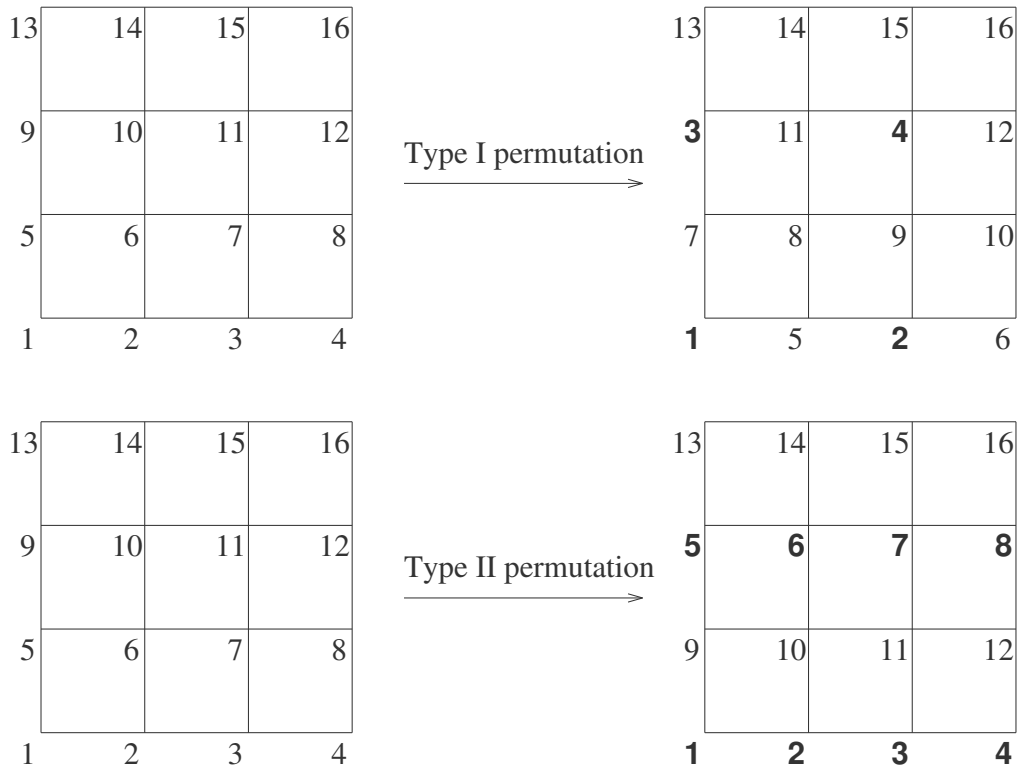}}
\caption{\footnotesize Example of a site permutation on a 4x4 lattice. Coarser lattice sites are labelled by boldface font.}
\label{lattice}
\end{figure}

We test the idea using the QCDLAB tool \cite{QCDLAB} and the new functions, {\tt Permutation\_block}, and {\tt Schur\_complement}. {\tt Permutation\_block} returns a permutation operator $P(p)$, which exchanges the rows of $D$, whereas {\tt Schur\_complement} is used to compute $S_{bb}$. The permutation $p$ is such that the coarser lattice sites are labelled first, as is shown in Figure \ref{lattice}.

We consider two types of permutation:
\begin{itemize}
\item Type I permutation: {\it block the lattice in all directions as shown in the upper panel of Figure 1.}
\item Type II permutation: {\it block the lattice in all directions, one direction at a time, as shown in the lower panel of Figure 1.}
\end{itemize}
Since to every permutation there is an inverse permutation, we have in all four types of permutations, $p_I,p_{II},p_I^{-1},p_{II}^{-1}$, the corresponding permutation operators being $P_I,P_{II},P_I^T,P_{II}^T$.

\begin{figure}
{\hspace{-2cm}\includegraphics[width=20cm,height=14cm]{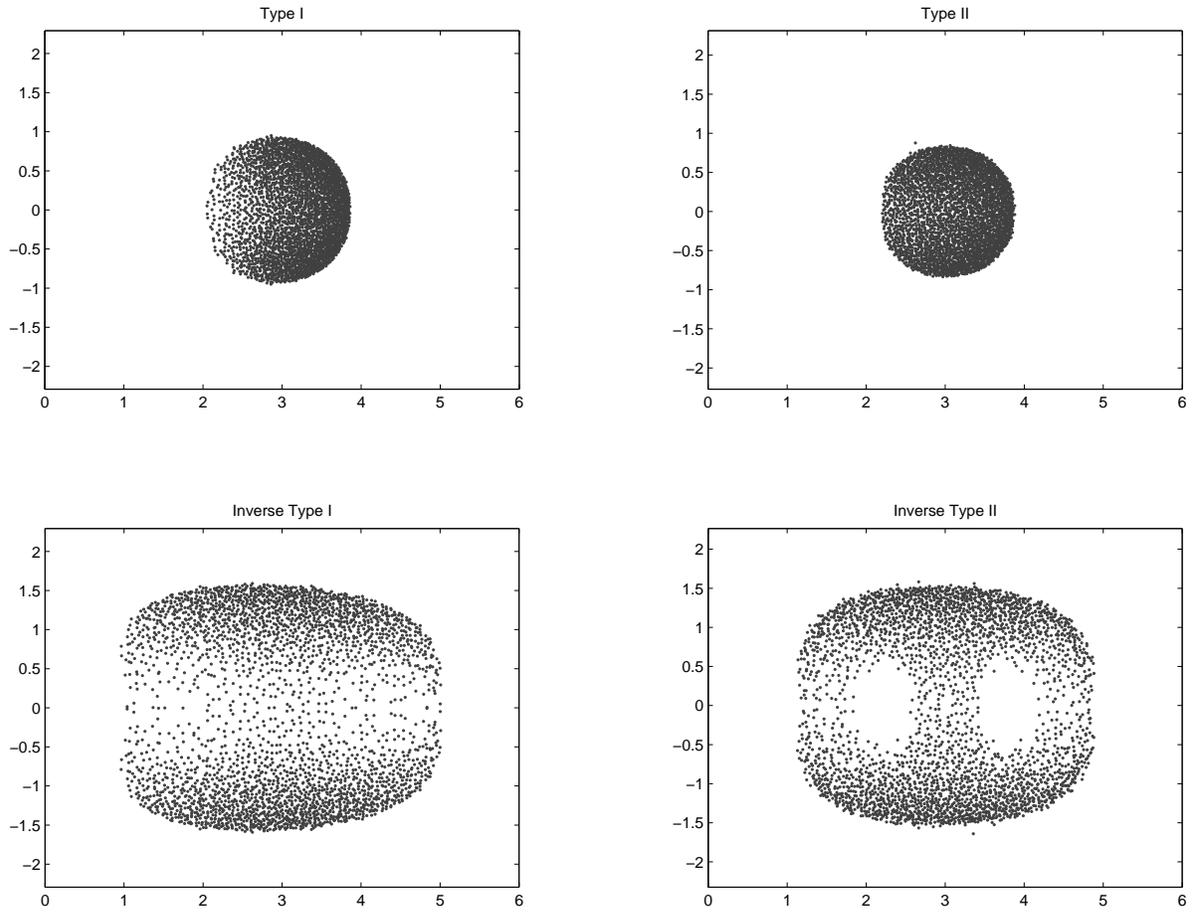}}
\caption{\footnotesize The spectrum of the blocked Wilson-Dirac operator using various permutations of lattice sites as described in the titles of the plots.}
\label{schur_complement}
\end{figure}

Using these permutations, we have computed four different Schur complements in a SU(3) background field using the plaquette action at $\beta=5.4$ on a $8^8$ lattice. At this coupling, the spectrum of the Wilson operator should have no spectral gap in the real axes \cite{Edwards_et_al98}. In Figure \ref{schur_complement} we show the effect of one step renormalisation group transformation on the spectrum of the Wilson-Dirac operator.

In the case of type II inverse permutation, one can see the emergence of an ``owl eyes'' plot. If such a picture is generic for this particular permutation, we could define the shifted Schur complement to be the domain wall or overlap kernel: if $c$ is the centre of the right ``eye'', the kernel of the chiral operator will be $cI-S_{bb}$. In this case, the benefits are threefold:
\begin{itemize}
\item[1.] better chiral properties/localisation of the domain wall/overlap operator;
\item[2.] faster inversions of the kernel;
\item[3.] better scaling properties of the lattice Dirac operator.
\end{itemize}

However, this construction could be computationally expensive: in order to compute the Schur complement one has to invert the $D_{rr}$ matrix. In fact, as it will be clear below, a Schur complement approximation to the exact one will do.

{\bf 3.} In principle, one can stay with the exact Schur complement if there is no need to iterate the renormalisation group transformation. One exmaple is the domain decomposition approach of L\"uscher \cite{luscher03}. He treats exactly the Schur complement and gets excellent results for the two-level algorithm. However, if one wants to iterate the scheme to full multigrid one has to rely on some approximation of the Schur complement. The reason is twofold: one would like to retain the sparsity of the operator and keep the computational overhead under control.

Earlier papers on Schur complement approximation have ignored the $D_{rr}$ matrix altogether \cite{wupp99}. Ignoring $D_{rr}$ has the advantage of conserving the sparsity pattern of the coarser operator. This is similar to Migdal-Kadanoff approximation, which according to Creutz, {\it ``its primary drawback lies in the difficulty of assessing the severity of the approximations involved''} \cite{Creutz_book}. On the other hand, an exact Schur complement may not be practical. The compromise is to allow the appearance of one power of $D_{rr}$.

\begin{figure}
{\hspace{2cm}\includegraphics[width=12cm,height=8cm]{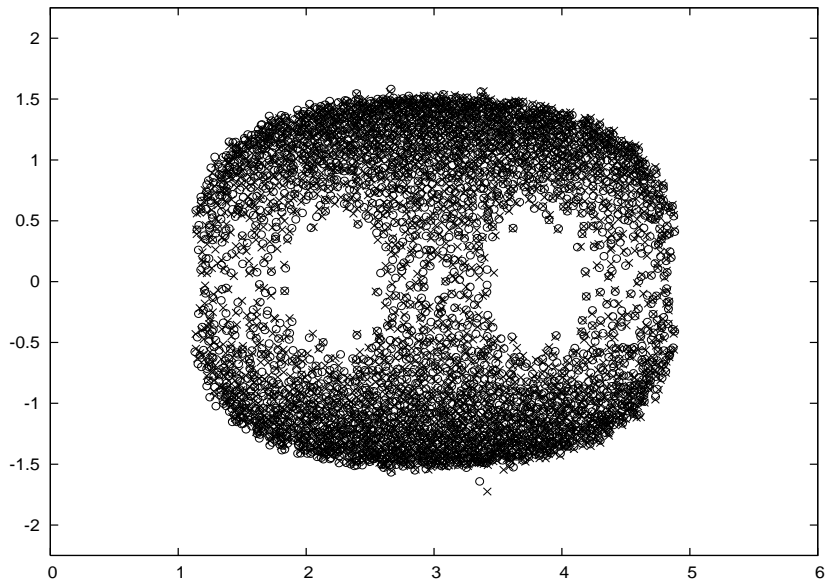}}
\caption{\footnotesize The spectrum of the Schur complement (circles) and its approximation (crosses).}
\label{schur_complement_appr_spectrum}
\end{figure}

The Schur complement approximation would be in the spirit of the renormalisation group transformation if the coarse lattice operator inherits basic properties of the fine lattice operator. In case of the Wilson operator one would like to conserve the positive definiteness, the covariance of the hopping matrix, and perhaps $\gamma_5$-Hermiticity. Reusken shows that employing point-Gaussian elimination on a weakly diagonally dominant M-matrix, which is similar to what we have, one gets a stable Schur complement approximation \cite{reusken00}. To the second order, his result reads:
$$
\tilde S_{bb}=D_{bb}-D_{br}\left(d_{rr}^{-1}+\tilde d_{rr}^{-1}-d_{rr}^{-1}D_{rr}\tilde d_{rr}^{-1}\right)D_{rb}\ ,
$$
where $d_{rr}$ is the diagonal of $D_{rr}$ and $\tilde d_{rr}$ is a diagonal matrix, its entries being the sum of $D_{rr}$ rows. Using this approximation, one can compare the properties of $S_{bb}$ and $\tilde S_{bb}$: the spectrum of $\tilde S_{bb}$ is again an ``owl eyes'' plot, as in Figure \ref{schur_complement_appr_spectrum}, whereas the cost of multiplication by $\tilde S_{bb}$ is approximately the same as the cost of multiplication by $D$.

\begin{figure}
{\vspace{-1cm}\hspace{-2cm}\includegraphics[width=20cm,height=14cm]{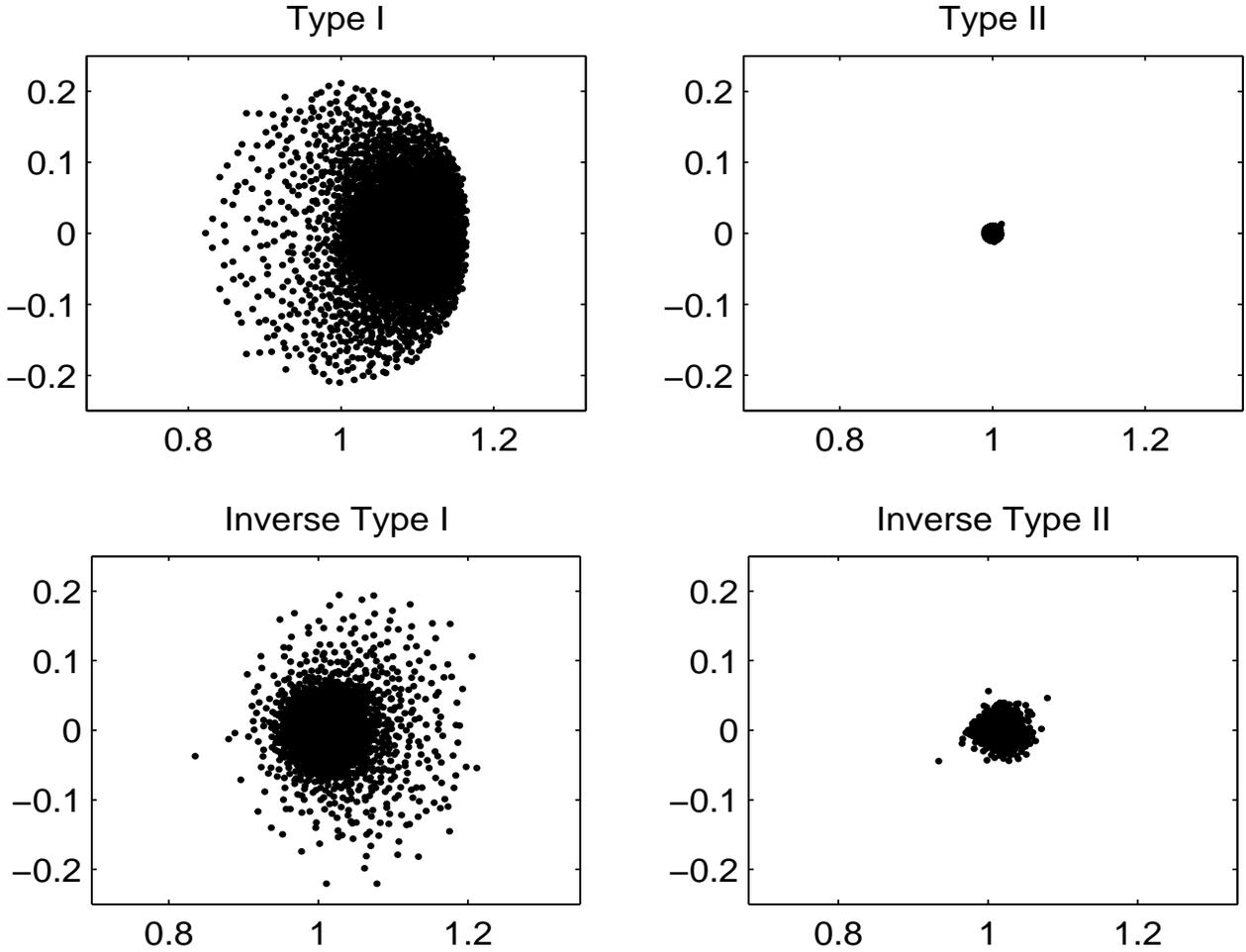}}
\vspace{-2cm}
\caption{\footnotesize The spectrum of $\tilde S_{bb}^{-1}S_{bb}$ using variuos permutations of lattice sites as described in the titles of the plots.}
\label{schur_complement_appr}
\end{figure}

{\bf 4.} In fact, if there is a place to test the Schur complement approximation, this is the full multigrid algorithm. Inheritance or stability is of great importance here, as well as regularity, i.e. how close an approximation is to the exact Schur complement. Without regularity, it would be impossible to iterate the two-grid algorithm to a full multigrid. From Figure \ref{schur_complement_appr} we see that, indeed, such an approximation gives a spectrum of $\tilde S_{bb}^{-1}S_{bb}$ clustered around one. Moreover, we observe that the type II and type II inverse permutations give excellent results.

Using such approximations of the Schur complement, one can construct a cyclic reduction preconditioner as in \cite{reusken00}. In our tests we used a 3-level V-cycle preconditioner: at the highest level, the Schur complement approximation is inverted exactly, whereas for the $D_{rr}$ inversion we used vanilla BiCGstab at one percent accuracy. We solved three linear systems to $10^{-8}$ accuracy at $\kappa=1/6$ on three $8^8$ lattices at $\beta=5.4,5.7,5.9$. The performance of the preconditioner is measured using the average convergence rate, $\cal R$, i.e. the average ratio of the next to the current residual error. In the second row of Table \ref{mg_results} we give the estimated condition numbers for each case. In the third and fourth row we show the results for the unpreconditioned and preconditioned BiCGstab algorithm. For the first two lattices we used the type I inverse permutation, whereas for the third one the type II inverse permutation. It should be noted that the preconditioned BiCGstab with the type I inverse permutation did not converge in the third case. According to \cite{luscher03}, the lattice at $\beta=5.9$ and bare quark mass $\kappa=0.1592$ corresponds to $m_{\pi}=320$ MeV, which indicates that our pion mass is close to its physical value. This shows that, even that close to criticality, a type II inverse permutation will yield a sufficient approximation to the Schur complement.

\begin{table}
\hspace{3cm}
\begin{tabular}{|l|l|l|l|}
\hline
$\beta$ & 5.4 & 5.7 & 5.9\\
\hline
Condition number &  57   &  480   &  905  \\
\hline
$\cal R$(BiCGstab) & 0.76   &  0.83   &  0.89  \\
\hline
$\cal R$(Preconditioned BiCGstab) & 0.05   &  0.16   &  0.48  \\
\hline
\end{tabular}
\vspace{0.5cm}
\caption{\footnotesize Performance of the 3-level V-cycle preconditioner on $8^8$ lattices for various couplings.}
\label{mg_results}
\end{table}

Applications of this preconditioner in lattice QCD with Wilson fermions are obvious:
\begin{itemize}
\item Acceleration of linear solvers: we didn't make any effort to construct a cheap preconditioner or to optimise its parameters; our prime interest was to test its potential. A production code should take this into account.
\item Acceleration of simulation algorithms: as it is already shown for DD-HMC algorithm \cite{luscher04}, the fermion determinant factorisation using the exact Schur complement can effectively be employed to accelerate the simulation algorithm. A full multigrid preconditioner offers the flexibility to employ more than two factors.
\end{itemize}

\end{document}